\documentclass[aps,prl,reprint,superscriptaddress]{revtex4-2}

\usepackage{amsmath}
\usepackage{amsfonts}
\usepackage{amssymb}
\usepackage{mathrsfs}
\usepackage{color}

\newcommand{\be}{\begin{equation}}
\newcommand{\ee}{\end{equation}}
\newcommand{\lf}{\widetilde{f}}
\newcommand{\cD}{\mathcal{D}}
\newcommand{\sD}{\mathcal{D}_x}
\newcommand{\tD}{\mathbb{D}}

\newcommand{\e}{{\rm e}}
\newcommand{\mL}{\mathcal{L}}
\newcommand{\mF}{\mathcal{F}}
\newcommand{\CC}{\mathbb{C}}
\newcommand{\II}{\mathbb{I}}
\newcommand{\RR}{\mathbb{R}}
\newcommand{\NN}{\mathbb{N}}

\newcommand{\lfrho}{\widehat{\widetilde{\rho}}}
\newcommand{\lW}{\widetilde{W}}
\newcommand{\lPhi}{\widetilde{\Phi}}

\newcommand{\lpsi}{\widetilde{\psi}}
\newcommand{\lPsi}{\widetilde{\Psi}}
\newcommand{\lK}{\widetilde{K}}

\begin{document}


\title{Master equations for continuous-time random walks 
with stochastic resetting}


\author{Fausto Colantoni}
\affiliation{Department of Basic and Applied Sciences for Engineering, 
Sapienza University of Rome, Via A. Scarpa 10, 00161 Rome, Italy}

\author{Gianni Pagnini}
\affiliation{BCAM--Basque Center for Applied Mathematics, 
Alameda de Mazarredo 14, 48009 
Bilbao, Basque Country -- Spain}
\affiliation{Ikerbasque--Basque Foundation for Science,
Plaza Euskadi 5, 48009 
Bilbao, Basque Country -- Spain}


\date{\today}

\begin{abstract}
We study a general continuous-time random walk (CTRW), 
by including non-Markovian cases and L\'evy flights,
under complete stochastic resetting to the initial position with
an arbitrary law, which can be power-lawed as well as Poissonian.
We provide three linked results.
First, we show that the random walk under 
stochastic resetting is a CTRW with the same jump-size distribution
of the non-reset original CTRW but different counting process. 
Later, we derive the condition for a CTRW 
with stochastic resetting to be 
a meaningful displacement process at large elapsed times, i.e., 
the probability to jump to any site is higher than 
the probability to be reset to the initial position,
and we call this condition {\it the zero-law for stochastic resetting}.
This law joins with the other two laws for reset random walks 
concerning the existence and the non-existence of a  
non-equilibrium stationary state.
Finally, we derive master equations for CTRWs 
when the resetting law is a completely monotone function.
\end{abstract}


\maketitle

Stochastic resetting is a topic of high interest 
in non-equilibrium statistical mechanics,
with a fast and large expansion during the last decade 
\cite{ctrw4,evansreview,kundu_etal-jpa-2024}.
Despite the vast amount of literature,
there are still theoretical open problems 
\cite{evans_etal-prl-2025,biroli_etal-pre-2025,liang_etal-pf-2025}. 
A non-exhaustive brief literature summary is the following.
Stochastic resetting is used for modelling diffusive phenomena 
that are forced randomly in time to re-start from a state which, 
in the simplest case, is the initial location 
\cite{evans2011diffusion,majumdar_etal-pre-2015,evansreview}.
More general resettings have also been studied \cite{dahlenburg_etal-pre-2021}. 
This mechanism can strongly affect the behaviour of the system
and lead to nontrivial non-equilibrium stationary state (NESS) 
\cite{evansreview,intermittent,anom1,ctrw-resetting-markovian}
or unsteady states \cite{powerlaw,shkilev_etal-jpa-2022}. 
The random interval between two resetting events 
can follow different types of distributions.
These distributions may be Poissonian (exponential) 
\cite{evans2011diffusion,majumdar_etal-pre-2015}, 
deterministic \cite{eliazar_etal-jpa-2020,eliazar_etal-jpa-2021}, 
which is often the most effective choice in search problems 
\cite{deterministic}, 
power-law \cite{powerlaw,ctrw-resetting-powerlaw,shkilev_etal-jpa-2022}, 
or follow other, more complex, forms \cite{intermittent}.
A large number of stochastic processes have been studied under 
the effect of resetting including 
normal diffusion \cite{evansreview}, 
L\'evy flights \cite{levy1,levy2}, 
scaled Brownian motion \cite{sbm1,sbm2}, 
and sub-diffusive anomalous diffusion 
\cite{anom1,ctrw-resetting-powerlaw,shkilev_etal-jpa-2022}. 
Since resetting introduces jumps and alters the dynamics of 
the underlying process, 
it is connected to fractional calculus and non-local operators 
especially in the time-reversed formulation \cite{boncoldovpag,coldovpag}. 
Stochastic resetting has proven to be advantageous in several contexts:
it can reduce the mean-first passage time in search problems 
\cite{chen_etal-pre-2022,linn_etal-pre-2023,mendez_etal-pre-2024}
and induce a stationary state in 
scale-free original processes \cite{anom1}.
Physically, it is related to
violation of microreversibility \cite{pal_etal-pre-2017} 
or to induced non-ergodicity and its restoring 
\cite{stojkoski_etal-pre-2021,stojkoski_etal-jpa-2022,
vinod_etal-pre-2022,wang_etal-prr-2022}. 
Beyond its theoretical interest, 
resetting has found applications in various fields 
including, for example, in computer science \cite{cs1,cs2},
biology \cite{bio1}, 
economy \cite{stojkoski_etal-prsa-2022,jolakoski_etal-csf-2023}.

We consider here stochastic resetting with arbitrary resetting rules
for continuous-time random walk (CTRW) \cite{montrollweiss65} 
in the most general case, 
that is by including both non-Markovianity and L\'evy flights.
The present formulation extends those considered in literature
\cite{anom1,ctrw-resetting-markovian,powerlaw,
ctrw-resetting-powerlaw,kusmierz_etal-pre-2019,shkilev_etal-jpa-2022}.
This model formulation has already been largely investigated in literature
\cite{ctrw1,ctrw2,ctrw3,ctrw4,ctrw5,
ctrw-resetting-markovian,ctrw-resetting-powerlaw,ctrw6} 
together with its extensions to network structures
\cite{netw1,netw2}. 
In the following, we study the effect of stochastic resetting  
to the initial position with a general law.
We provide {\it three} linked results.
{\it First}, we show that a random walk under 
stochastic resetting can be represented as a CTRW without resetting. 
Thus, any random walk under stochastic resetting is indeed
a special non-reset random walk.
{\it Later}, we derive the condition for a CTRW 
under stochastic resetting to be 
a meaningful displacement process, 
that is a process whose trajectories diffuse in time, 
{\it saltem aliquamdiu}, 
and they are not dominated by the resetting to the starting location.
Because of its basic role for the existence of a meaningful system with reset,
we call this condition {\it the zero-law for stochastic resetting}.
This law joins with the other two laws for reset random walks 
concerning the existence and the non-existence of a NESS \cite{evansreview}.
The derived zero-law may be applied for inferring the family of the
underlying non-reset random walk provided that 
a non-stuck reset process is observed  
and the distribution of the resetting intervals is known. 
{\it Finally}, we derive the master equation for the random walk
under resetting when the resetting law is a completely monotone function. 

Before to start with technicalities,
we report that mathematical notions and notations used in the following
are given in the End Matter section.

We denote by $X_t$ and $Y_t$ the trajectory of the diffusing particle
and of the reset-process that are characterized 
by the corresponding probability density function (p.d.f.)
$\rho(x;t)$ and $p(x;t)$, respectively, 
for the space variable \(x \in \mathbb{R}\) and time \(t>0\).
In particular,
the trajectory $X_t$ is generated by an uncoupled one-dimensional CTRW 
\cite{montrollweiss65} where the i.i.d. waiting-times,
with law $\psi$,
and the i.i.d. jump-size distributed, with law $\lambda$, 
are, at any epoch, random variables independent of each other 
and also of the current position and time, 
such that their joint-law factorizes in term of the marginal distributions. 
Thus, at any random times \(t_i > 0\), $i \in \NN$,
from the random jump-sizes \(\xi_i = X_{t_i} - X_{t_{i-1}}\)
and the waiting-times between 
two consecutive jumps \(\tau_i=t_i - t_{i-1}\) 
we have
	 \[ X_t:=\sum_{i=1}^{N_t} \xi_i \,,
	 \]
where \(t_0=0\), \(X_0=0\), and \(N_t\) is a counting process
that provides the random number of jumps up to time \(t\),
such that the probability to have an elapsed time equals to $t$ after $n$ jumps 
is here denoted by \(\mathbf{P}(N_t=n)= P_n(t)\) and is defined by
\[
P_n(t)=\int_0^t P_{n-1}(t-\tau) \psi(\tau) \, d\tau \,,
\quad P_0(t)=\Psi(t) \,,
\]
where \(\Psi(t) = 1 - \int_0^t \psi(\tau) \, d\tau\) is called
survival probability.
In this framework,
it holds \cite[Formula (16)]{scalas-schilling-ctrw}
\be
\label{rho}
\rho(x;t)= \sum_{n=0}^\infty \rho_n(x) P_n(t) \,,
\ee
 where \(\rho_n(x)\) is the discrete-time walker's p.d.f. 
to be in \(x\) after $n$ steps, 
which corresponds to the \(n-\)fold convolution 
of the marginal probability density of the jump-sizes \(\lambda\). 
Moreover, from \eqref{rho}, 
in the Fourier--Laplace domain the p.d.f. $\rho(x;t)$ results
in the Montroll--Weiss formula \cite{montrollweiss65}
\be
\label{rho-transform}
\lfrho(k;s)=\lPsi(s) \sum_{n=0}^\infty 
[\widehat{\lambda}(\kappa)]^n [\lpsi(s)]^n
=\frac{\lPsi(s)}{1- \widehat{\lambda}(\kappa) \lpsi(s)} \,.
\ee
L\'evy flights are included,
then jump-size distribution $\lambda$ can display power-law tails,
i.e., $\lambda(x) \sim |x|^{-\alpha - 1}$ when $|x| \to \infty$
with $0 < \alpha < 2$, together with classical exponential or Gaussian tails.
Non-Markovianity also is included,
then the waiting-time distribution is assumed to have
Laplace transform \cite{hilfer_etal-pre-1995}
\be
\lpsi(s)=\frac{1}{s^\beta+1} \,, 
\quad \lPsi(s)=\frac{s^{\beta-1}}{s^\beta+1} \,,
\quad 0 < \beta < 1 \,.
\label{WTnonM}
\ee 
Thus, the master equation of the process $X_t$ is
\cite{mainardi_etal-fcaa-2001}
\be
\cD^\beta \rho=\sD^\alpha \rho \,, \quad \rho(x;0)=\delta(x) \,,
\ee
or, analogously,
\be
\frac{\partial \rho}{\partial t}
=\tD^{1-\beta} \sD^\alpha \rho \,, \quad \rho(x;0)=\delta(x) \,,
\ee
see the End Matter section for the meaning of $\cD^\beta$, 
$\tD^{1-\beta}$ and $\sD^\alpha$.
	 
We now introduce the law of the 
complete stochastic resetting to the initial position, 
which in our case is set equal to zero. 
Let \(T_i\) , with $i \in \NN$, 
be the sequence of the i.i.d. resetting intervals
with law \(\varphi\):
\[
\mathbf{P}(T_i \in dt)= \varphi(t) \, dt \,, \quad \forall i \in \NN \,.
\]
The resetting points are defined by
\(S_k= \sum_{i=1}^k T_i\), with $k=1,2,...$,
and the renewal process is
\(M_t= \sum_{k=1}^{\infty} \mathbf{1}(S_k \leq t) \).
We consider a broad class of resetting-intervals distributions,
which allows for generalizing known cases with 
power- \cite{ctrw-resetting-powerlaw,shkilev_etal-jpa-2022}
and Poissonian-law 
\cite{evans2011diffusion,evansreview,ctrw-resetting-markovian}. 

\smallskip	 
{\it Any CTRW under resetting is another CTRW---}By following literature \cite[Section 2.8]{evansreview}, 
we use a renewal approach to determine the p.d.f. \(p(x;t)\). 
Let \(\Phi(t)\) be the probability of no resets up to time \(t\)
defined by
	 \[\Phi(t)= 1 - \int_0^t \varphi(\tau) \, d\tau \,,\]
therefore the distribution of no resetting in \([0, t]\) is  
	 \[\mathbf{P}_0(Y_t \in dx \vert \text{ no reset in } [0,t])
= \Phi(t)\rho(x;t) \, dx \,, \]
	 because with no resetting \(Y_t=X_t\). 
If we have one reset \(S_1 \in [0,t]\), we write
\begin{eqnarray}
&&\mathbf{P}_0\left(Y_t \in dx \vert \text{ one reset in } [0,t]\right) 
= \qquad \qquad \nonumber \\
&& \qquad \qquad 
\int_0^t \varphi(S_1) \Phi(t - S_1) \rho(x;t- S_1) \, dS_1 dx \,, 
\end{eqnarray}
where \(\varphi\) governs the probability that one reset event happens 
in \(S_1\),
after which the process restarts from zero with the law \(\rho(x;t)\), 
and by $\Phi(t - S_1)$ we satisfy the request 
that there are no any other resets in the interval \((S_1, t]\). 
Similarly for two resets in \(S_1, S_2\), we get
\begin{eqnarray}
&& \mathbf{P}_0\left(Y_t \in dx \vert \text{ two resets in } [0,t]\right) =
\qquad \qquad \nonumber \\
&& \qquad \qquad \int_0^t 
\int_0^{t-S_1} \varphi(S_1) \varphi(S_2) \Phi(t - S_1 - S_2)
\nonumber \\
&& \qquad \qquad \qquad \qquad \times \rho(x;t-S_1-S_2) \, dS_2 dS_1 dx \,.
\end{eqnarray}
Since we are handling convolution integrals, 
it is convenient to pass to the Laplace domain and, 
by applying the same idea recursively for \(n\) resets, we obtain
	 \begin{align*}
	 	\widetilde{p}(x;s) \, dx 
&= \int_0^\infty \e^{-s t} \mathbf{P}_0(Y_t \in dx) \, dt \\
	 	&= \int_0^\infty \sum_{n=0}^\infty \e^{-st} 
\mathbf{P}_0\left(Y_t \in dx \vert \, n \text{ resets in } [0,t] \right) \, dt\\
	 	&= \sum_{n=0}^\infty [\widetilde{\varphi}(s)]^n 
 \int_0^\infty \e^{-st} \Phi(t) \rho(x;t) \, dtdx\\
	 	&=\frac{1}{1-\widetilde{\varphi}(s)} 
\int_0^\infty \e^{-st} \Phi(t) \rho(x;t) \, dtdx \,,
	 \end{align*} 
and then
	 \begin{align}
	 	\label{p-lap}
	 	\widetilde{p}(x;s)
= \frac{1}{s \widetilde{\Phi}(s)} \int_0^\infty \e^{-st} \Phi(t) \rho(x;t) 
\, dt \,.
	 \end{align}
After the invertion of the Laplace transform, we have
	 \begin{align}
	 	\label{p-density}
p(x;t) = \Phi(t) \rho(x;t) + \int_{0}^t \varphi(\tau) p(x;t-\tau) \, d\tau \,.
	 \end{align}
Formula \eqref{p-density} holds for  
non-Markovian processes with non-Poissonian resettings
by generalizing existing literature results. 
In particular, when process $X_t$ is Markovian the semigroup formalism
can be used, namely $\rho(x;t)=\e^{t \mathcal{L}(x)}\rho(x;0)$,
and when the resetting is Poissonian $\Phi(t)=\varphi(t)=\e^{-t}$, 
such that, in each case, Laplace transform \eqref{p-lap} reduces 
to a known formula \cite{intermittent,kusmierz_etal-pre-2019}.
Expressions derived in equations \eqref{p-lap} and \eqref{p-density} 
are completely general, without assuming that \(X_t\) is a CTRW. 
When $X_t$ is assumed to be a CTRW, then
the process \(Y_t\) with density \(p(x;t)\) given by \eqref{p-density} 
is a CTRW in the form
	 	\begin{align}
	 		\label{p}
	 		p(x;t)= \sum_{n=0}^\infty \rho_n(x) \overline{P}_n(t) \,,
	 	\end{align}
where \(\rho_n\) is the same introduced in \eqref{rho} and \(\overline{P}_n\) 
is related to \({P}_n\) via the Volterra integral equation
\begin{align}
\label{PsegnatoP}
\overline{P}_n(t)  = 
P_n(t) \Phi(t) + \int_0^t \varphi(\tau) \overline{P}_n(t-\tau) \, d\tau \,.
\end{align}
Then, \(Y_t\) is a CTRW with the same jump-size distribution of \(X_t\) 
and the following representation holds
\[Y_t:=\sum_{i=1}^{Q_t} \xi_i \,,
\quad 
Q_t = N_t - \sum_{k=1}^{M_t} Q_{S_k} \,,
\]
with \(\mathbf{P}(Q_t =n) = \overline{P}_n(t)\). 
We note that the distributions conditioned on the number of resets, 
which were used to derive equations \eqref{p-lap} and \eqref{p-density}, 
do not affect the jump-sizes, 
but act only on the temporal component. 
In fact, we are applying the resetting mechanism to the process \( N_t \), 
which evolves according to the distribution \( P_n \). 
Thus, the new counting process \(Q_t\)
with law \(\overline{P}_n\) 
derives from the resetting mechanism on the process \(N_t\).

\smallskip
{\it Zero-law of stochastic resetting---}From \eqref{PsegnatoP}, 
we have that the Laplace transform of $\overline{P}_n(t)$ is
\be
\widetilde{\overline{P}_n}(s)
=\frac{1}{s \widetilde{\Phi}(s)} \int_0^\infty e^{-st} \Phi(t) P_n(t) dt \,,
\label{psegnato-lap}
\ee
and since both \(P_n\) and \(\Phi\) are probability, 
i.e., both are \(\leq 1\), then we have the estimate 
\begin{eqnarray}
\widetilde{\overline{P}_n}(s)
&=&\frac{1}{s \widetilde{\Phi}(s)} \int_0^\infty e^{-st} \Phi(t) P_n(t) \, dt 
\nonumber \\
&\leq& \frac{1}{s \widetilde{\Phi}(s)} \int_0^\infty e^{-st}  P_n(t) \, dt 
=\frac{\widetilde{\Psi}(s) 
\left[\widetilde{\psi}(s)\right]^n}{s \widetilde{\Phi}(s)} \,. \quad 
\end{eqnarray}
Moreover, we know that 
$\overline{P}_n(t) \leq \overline{P}_0(0) = 1$, 
which, in the Laplace domain, reads 
$\widetilde{\overline{P}_n}(s) \leq \frac{1}{s}$.
We observe that, analogously, 
\[\lpsi(s)= \int_0^\infty e^{-st} \psi(t) \, dt 
\leq \int_{0}^{\infty} \psi(t) dt = 1 \,, \]
therefore, 
\(\left[\widetilde{\psi}(s)\right]^n \to 0\) for \(n \to \infty\) and \(s>0\). 
Thus, the following limit holds
\[
\widetilde{\overline{P}_n}(s)=
\frac{\lPsi(s) \left[\lpsi(s)\right]^n}{s \widetilde{\Phi}(s)} \to 0 \,,
\quad n \to \infty \,, \quad s>0 \,.
\]
Hence, we can conclude that there exists an index \( n_0 \) 
such that the following inequality is true for all \( n > n_0 \)
\be
\widetilde{\overline{P}_n}(s)=
\frac{\widetilde{\Psi}(s) \left[\widetilde{\psi}(s)\right]^n}
{s \widetilde{\Phi}(s)} \leq \frac{1}{s} \,, \quad \forall n > n_0 \,.
\label{Pninequality}
\ee
Since \( \overline{P}_n \) is a probability, 
we are interested only in functions that satisfy \eqref{Pninequality} 
because its maximum must be equal to $1$.
For our purposes, we rewrite \eqref{Pninequality} as
\be
\widetilde{\Psi}(s) \left[\widetilde{\psi}(s)\right]^n 
\leq \widetilde{\Phi}(s) \,, \quad \forall n > n_0 \,,
\label{eq:ineq}
\ee
and, actually, it can be turned into
\be
1 - \widetilde{\Psi}(s) \left[\widetilde{\psi}(s)\right]^n 
\geq 1 - \widetilde{\Phi}(s) \,, \quad \forall n > n_0 \,,
\ee
which means that, 
after $n$ iterations since the beginning, 
the probability to jump towards any site 
is larger than the probability to be reset towards the 
starting site $x=0$. Thus, 
at large elapsed times,
the reset CTRW is a meaningful displacement process 
whose trajectories diffuse in time 
notwithstanding stochastic resetting.
From \eqref{eq:ineq} it follows
\[
(1-\lpsi(s)) \left[\lpsi(s)\right]^n \leq (1-\widetilde{\varphi}(s)) \,, \]
and finally, 
since \(\log \widetilde{\psi}(s) <0\), 
we have a condition for \(n\)
\be
n \geq \frac{\log(1-\widetilde{\varphi}(s)) 
- \log(1-\widetilde{\psi}(s))}{\log \widetilde{\psi}(s)} \,,
\label{condn}
\ee
which provides the number of jumps \(n\) required 
to observe a jump-dominated process generated 
by a CTRW with waiting-time distribution 
\(\psi\) and resetting-interterval distribution \(\varphi\).
We call condition \eqref{condn} 
the {\it zero-law of stochastic resetting}
because it provides the condition for a meaningful 
displacement process under reset.
Heuristically, 
condition \eqref{condn} determines a kind of time-scale in terms of 
the number of jumps necessary to put in motion the mechanism 
of jumps and resettings, 
namely $n_0$, which is reflected into a frequency-scale, 
say $s_0$ with $n_0 \, s_0 = \mathcal{O}(1)$.
Thus, this time-scale defines the small-time behaviour ($s \gg s_0)$
dominated by the resetting and 
the large-time behaviour ($s \ll s_0$) 
dominated by the diffusion.
In other words, 
condition \eqref{condn} is a criterium for the existence of 
a large-time behaviour with respect to $n_0$ where
a displacement-regime driven by jumps over resettings is displayed.
In particolar, we have the following.

Since $n \ge 0$, condition \eqref{condn} is
met whenever $\log(1-\widetilde{\varphi}(s))
- \log(1-\widetilde{\psi}(s)) > 0$ because $\log \widetilde{\psi}(s) < 0$.
Then, condition \eqref{condn} is always satisfied 
if $\widetilde{\varphi}(s) < \widetilde{\psi}(s)$.
By taking a resetting with a power-law,
say $\widetilde{\varphi}=1/(s^\mu + 1)$, with $0 < \mu < 1$,
the reset process turns out to be always meaningful if 
$\widetilde{\varphi}(s) < \widetilde{\psi}(s)$ provided that
$s^\mu > s^\beta$ when $s \to 0$, that means $\mu < \beta$:
which includes the case of the Brownian motion ($\beta=1$) 
under power-law ($0 < \mu < 1$) \cite{powerlaw}.

Actually, 
when $\widetilde{\varphi}(s) > \widetilde{\psi}(s)$, condition \eqref{condn} 
transforms into a further condition for $\widetilde{\varphi}$ and $\lpsi$
that is
\be
\widetilde{\varphi} \ge 1 - \lpsi^n(1-\lpsi) \,,
\quad s \le s_0 \,. 
\label{zerolaw}
\ee
Condition \eqref{zerolaw} holds in inequality when $s < s_0$,
in equality when $s=s_0$ and it does not hold when $s > s_0$.
In particular, we have that $n=0$
when $\widetilde{\varphi}=\lpsi$ and,
in opposition, the reset process is never observed
($n=\infty$)
when $\lpsi(s)=1$ except if also $\widetilde{\varphi}=1$. 
On the other side, if $\widetilde{\varphi}=1$
then any $\lpsi(s)$ is valid.
By taking again a resetting with a power-law,
the reset process is meaningful if
$\widetilde{\varphi}(s) > \widetilde{\psi}(s)$,
provided that $s^\beta (s^\mu+1) \ge s^\mu(s^\beta+1)^{n+1}$ 
when $s \to 0$, that means $\mu > \beta$:
which includes the case of a non-Markovian CTRW ($0 < \beta < 1$) 
under Poissonian resetting ($\mu=1$) 
\cite{anom1,ctrw-resetting-markovian}.

The {\it zero-law of stochastic resetting} \eqref{condn}
is the complement and the extension based on probabilistic argouments
of a similar schematizing analysis in literature 
\cite{ctrw-resetting-powerlaw}. 

{\it Master equations for CTRWs with reset---}We conclude our study deriving the master equation for
the processes with resetting. 
Provided that condition \eqref{condn} is met.
Our findings extend literature results 
\cite{intermittent,kusmierz_etal-pre-2019}.
We assume that the i.i.d. intervals 
between resettings \(T_1, T_2,...\) have a completely monotone law
\cite{schilling-book}
\begin{align}
\label{cm}
\varphi(t)= \int_0^\infty \e^{-rt} \, F(dr) 
= \int_0^\infty  r \e^{-rt} f(r) \, dr \,,
\end{align}
where, in general, we consider \(F\) absolutely continuous 
with non-negative density \(f\) such that 
$\int_0^\infty f(r) \, dr =1$. 
Distributions \(\varphi\) are \textit{completely monotone}
because they are the Laplace transform of a non-negative function 
\cite[Theorem 1.4]{schilling-book}. 
It can be checked that the Poissonian exponential distribution 
\(\theta \, \e^{-\theta t} \), for \(\theta >0\), belongs
to this family of functions. 

Since we have seen that \(p(x;t)\) is the p.d.f. of a CTRW \eqref{p}, 
we look for its Montroll--Weiss formula in analogy with \eqref{rho-transform}, 
i.e.,
\be
\widehat{\widetilde{p}}(k;s)= 
\frac{\widetilde{W}(s)}{1- \widehat{\lambda}(k) \widetilde{w}(s)} \,,
\ee
where 
$\widetilde{W}(s)=(1 - \widetilde{w(s)})/s$ and $w$ 
is the law of the waiting-times of the reset CTRW \(Y_t\). 
We recall that the jump-size distribution 
is the same for both \(X_t\) and \(Y_t\), 
then from \eqref{p}, we have
\[\widehat{\widetilde{p}}(k;s) 
= \sum_{n=0}^\infty [\widehat{\lambda}(k) ]^n \widetilde{\overline{P}_n}(t) 
\,.
\] 
We observe that when \(\varphi\) is a complete monotone distribution
according to \eqref{cm}, it results
\[\Phi(t)= 1- \int_0^t\varphi(y) \, dy=\int_0^\infty \e^{-yt} f(y) \, dy\]
and
\[\widetilde{\Phi}(s)= \int_{0}^{\infty} \frac{f(y)}{s+y} \, dy \,.\]
Now, to obtain the Montroll-Weiss type formula, 
we look for a Laplace transform such that
\be
\widetilde{\overline{P}_n}(s)= \widetilde{W}(s) \widetilde{w}_n(s) 
= \frac{\widetilde{w}_n(s) - \widetilde{w}_{n+1}(s)}{s} \,,
\ee
and it holds for
\be
\widetilde{w}_n(s)=
\frac{\int_0^\infty [\widetilde{\psi}(s+y)]^n \frac{f(y)}{s+y} dy}
{\int_{0}^{\infty} \frac{f(y)}{s+y} \, dy} \,,
\ee
where we used the notation \(\widetilde{w}(s)=\widetilde{w}_1(s)\).
In fact, from \eqref{psegnato-lap}, we have
\begin{align*}
	\widetilde{\overline{P}_n}(s)
&=\frac{1}{s \widetilde{\Phi}(s)} 
\int_0^\infty \e^{-st} \Phi(t) P_n(t) \, dt\\
&=\frac{1}{s \widetilde{\Phi}(s)} 
\int_0^\infty \int_0^\infty \e^{-t(s+y)} P_n(t) f(y) \, dt dy\\
	&= \frac{1}{s \widetilde{\Phi}(s)} 
\int_0^\infty \widetilde{\Psi}(s+y) [\widetilde{\psi}(s+y)]^n f(y) \, dy\\
	&=\frac{1}{s \widetilde{\Phi}(s)} 
\int_0^\infty \frac{1-\widetilde{\psi}(s+y)}{s+y} 
[\widetilde{\psi}(s+y)]^n f(y) \, dy\\
	&=\frac{1}{s \widetilde{\Phi}(s)} 
\int_0^\infty ( [\widetilde{\psi}(s+y)]^n- [\widetilde{\psi}(s+y)]^{n+1}) 
\frac{f(y)}{s+y} \, dy \,.
\end{align*}

Then the Fourier--Laplace transform of the p.d.f. $p(x;t)$ is
\begin{eqnarray}
\widehat{\widetilde{p}}(k;s) 
&=&\widetilde{W}(s)\sum_{n=0}^\infty [\lambda(k)]^n \widetilde{w}_n(s)
\nonumber \\
&=&\widetilde{W}(s) \frac{\int_0^\infty \sum_{n=0}^\infty 
[\widehat{\lambda}(k)]^n 
[\widetilde{\psi}(s+y)]^n \frac{f(y)}{s+y} \, dy}
{\int_{0}^{\infty} \frac{f(y)}{s+y} \, dy} \nonumber \\
&=&\lK(s) \int_0^\infty
\frac{\lfrho_*(\kappa;s+y)}{s+y} f(y) \, dy \,,
\label{lfp}
\end{eqnarray}
where 
\be
\lK=\frac{\lW}{\lPhi}=\frac{1}{s \lPhi^2}
\left[\lPhi - \int_0^\infty \frac{\lpsi(s+y)}{s+y}f(y) \, dy 
\right] \,,
\label{K}
\ee
and, by assuming the non-Markovian formulation \eqref{WTnonM},
\be
\lfrho_*(\kappa;s)
=\frac{\lfrho(\kappa;s)}{\lPsi(s)}
=(s^\beta+1) \, \frac{\lfrho(\kappa;s)}{s^{\beta-1}} \,,
\label{lfrhostar}
\ee
that, in the original domain, reads
\begin{eqnarray}
\rho_*(x;t)
&=& \frac{\partial \rho}{\partial t} + \cD^{1-\beta} \rho 
+ \delta(x)\left[\frac{t^{-(1-\beta)}}{\Gamma[1-(1-\beta)]} +
\delta(t) \right] \nonumber \\
&=& \frac{\partial \rho}{\partial t} + \tD^{1-\beta} \rho 
+ \delta(x)\delta(t) \,.
\label{rhostar}
\end{eqnarray}
Moreover, from \eqref{lfp} we have
(see, properties of Laplace transform reminded in the End Matter section)
\begin{eqnarray}
p(x;t)
&=& \int_0^t K(t-\tau)\int_0^\infty \left\{
\e^{-y\tau} \int_0^\tau \rho_*(x;\zeta) \, d\zeta \right\} f(y) \, dyd\tau
\nonumber \\
&=& \int_0^t K(t-\tau) \left\{
\int_0^\infty \e^{-y\tau} f(y) \, dy \right\}
\int_0^\tau \rho_*(x;\zeta) \, d\zeta d\tau
\nonumber \\
&=& \int_0^t K(t-\tau) \lf(\tau) 
\int_0^\tau \rho_*(x;\zeta) \, d\zeta d\tau \,,
\end{eqnarray} 
that, by using the first line of \eqref{rhostar}, gives
\be
p(x;t)
= \int_0^t K(t-\tau) \lf(\tau) 
\left[\rho(x;\tau) + J^\beta \rho \right] \, d\tau \,.
\label{int:p}
\ee
Therefore, by using the convolution representation of $p(x;t)$ \eqref{int:p},
we have
(see, the Caputo fractional derivative of a Laplace convolution integral
in the End Matter section)
\begin{eqnarray}
\cD^\beta p 
&=& \int_0^t K(t-\tau) \cD^\beta [\lf(\rho +J^\beta \rho)] \, d\tau
\nonumber \\
& & \quad 
+ \delta(x)\left[\lf(0)J^{1-\beta}K - \frac{t^{(1-\beta)-1}}{\Gamma(1-\beta)}
\right] \,.
\end{eqnarray}
We can also derive the following
\begin{eqnarray}
\sD^\alpha p 
&=& \int_0^t K(t-\tau)\lf(\tau)
\left[\sD^\alpha \rho + J^\beta \sD^\alpha \rho \right] \, d\tau \nonumber \\
&=& \int_0^t K(t-\tau)\lf(\tau)
\left[\rho - \delta(x) + \sD^\alpha \rho \right] \, d\tau \,,
\end{eqnarray}
where we used the equalities
$$
J^\beta \sD^\alpha \rho =
J^\beta \cD^\beta \rho =
J^\beta J^{1-\beta}D^1 \rho =
J^1 D^1 \rho = \rho - \delta(x) \,.
$$
Finally, master equations for CTRWs under stochastic resetting 
to the initial position are
\begin{eqnarray*}
\cD^\beta p 
&=& \sD^\alpha p  \nonumber \\
& & + \int_0^t K(t-\tau) \cD^\beta \left[
\lf(\tau) \left(\rho(x;\tau) + J^\beta \rho \right)\right] \, d\tau
\nonumber \\
& &  
- \int_0^t K(t-\tau) \lf(\tau)\left[
\rho(x;\tau) - \delta(x) + \sD^\alpha \rho \right] 
\, d\tau
\nonumber \\
& & + \delta(x) \left[ J^{1-\beta}K \lf(0) 
- \frac{t^{(1-\beta)-1}}{\Gamma(1-\beta)}\right] \,,
\quad p(x;0)=\delta(x) \,,
\end{eqnarray*}
provided that $\widetilde{\varphi}$ and $\lpsi$ met 
the {\it zero-law of stochastic resetting} \eqref{condn}.
We can derive the master equation in the Markovian case, i.e., $\beta=1$,
with Poissonian resetting, i.e., $f(y)=\delta(y-1)$,
such that $\widetilde{K}(s)=(s+1)^2/[s(s+2)]$ and 
it turns out to be
$$
\frac{\partial p}{\partial t} = \sD^\alpha p - p(x;t) + \delta(x) \,.
$$
By setting $\alpha=2$ we have the well-known equation
for Brownian motion under Poissonian resetting 
\cite{evans2011diffusion,evansreview,ctrw-resetting-markovian}. 

\smallskip
\begin{acknowledgments}
{\it Acknowledgments---}F.C. acknowledges Sapienza and the group INdAM-GNAMPA for their support
and thanks BCAM--Basque Center for Applied Mathematics, Bilbao, 
for the hospitality during a three-months visiting period. 
F.C. is funded by MUR under the project 
PRIN 2022--2022XZSAFN: Anomalous Phenomena on Regular and Irregular Domains:
Approximating Complexity for the Applied Sciences - CUP B53D23009540006.
Web Site: \url{https://www.sbai.uniroma1.it/~mirko.dovidio/prinSite/index.html}
G.P. is supported by the Basque Government through the BERC 2022--2025 program 
and by the Spanish Ministry of Science and Innovation: 
BCAM Severo Ochoa accreditation 
CEX2021-001142-S / MICIN / AEI / 10.13039/501100011033.
\end{acknowledgments}


%

\section*{End matter}
We report here the mathematical notions and notations  
used in the main text. 

Let $\mL\{g(t),s\}=\widetilde{g}(s)=\int_0^\infty \e^{-st}g(t) \, dt$,
with $s \in \CC$, 
be the Laplace transform
and $\mF\{u(x),\kappa\}=\widehat{u}(\kappa)=
\int_{-\infty}^{+\infty}\e^{+i\kappa x}u(x) \, dx$, with $\kappa \in \RR$,
be, in analogy with the definition of characteristic function, 
the Fourier transform of sufficiently well behaving functions 
$g(t)$ and $u(x)$.

Let $J^\sigma$, with $\sigma >0$, be the Riemann--Liouville fractional
integral of order $\sigma$ defined by \cite{gorenflo_etal-cism-1997}
$$
J^\sigma g(t)=\frac{1}{\Gamma(\sigma)}
\int_0^t \frac{g(\tau)}{(t-\tau)^{1-\sigma}} \, d\tau \,,
\quad t>0 \,,
$$
such that $J^0=\II$ and 
$\mL\{J^\sigma g\}=\widetilde{g}/s^\sigma$.
By using the Laplace transform, it can be shown that the
semigroup property holds, i.e.,
$J^\sigma J^\eta = J^{\sigma+\eta}$.
A noteworthy result used in our study is
$J^\sigma 1= t^\sigma/\Gamma(1+\sigma)$.

Moreover, let $\cD^\beta$ and $\tD^\beta$ be the
time-fractional derivative of order $\beta$
in the Caputo and the Riemann--Liouville sense,
respectively, defined by \cite{gorenflo_etal-cism-1997}
$$
\cD^\beta g = J^{m-\beta}D^m g \,, \quad
\tD^\beta g = D^m J^{m-\beta} g \,,
$$
with $m-1<\beta<m$, $m\in\NN$, $\cD^m=\tD^m=D^m$
and $\cD^0=\tD^0=D^0=\II$. 
The corresponding Laplace transforms are
$$
\mL\{\cD^\beta g,s\} = s^\beta \widetilde{g} 
- \sum_{\ell=0}^{m-1} g^{(\ell)}(0^+) s^{\beta-1-\ell} \,, 
\label{cLaplace}
$$
$$
\mL\{\tD^\beta g,s\} = s^\beta \widetilde{g} 
- \left. \sum_{\ell=0}^{m-1} D^\ell J^{m-\beta}g \right|_{0^+} s^{m-1-\ell} \,. 
\label{tLaplace}
$$
By using the above formula for $\mL\{\cD^\beta g,s\}$, 
the Caputo fractional derivative
of a Laplace convolution integral can be determined as follows.
Consider the Laplace convolution integral
$q(t)=\int_0^t g(t-\tau)h(\tau) \, d\tau$, 
then by using the rule $\widetilde{q}=\widetilde{g} \, \widetilde{h}$
and the Laplace transform of the Caputo fractional derivative,
i.e., $\mL\{\cD^\beta g,s\}$, 
we have
\begin{eqnarray*}
s^\beta \widetilde{q} - s^{\beta-1}q_0
&=& s^\beta \widetilde{g}\widetilde{h} - s^{\beta-1}q_0
\nonumber \\
&=& \widetilde{g} [s^\beta \widetilde{h} - s^{\beta-1}h_0]
+ s^{\beta-1} \widetilde{g}h_0 
- s^{\beta-1} q_0 \,,
\end{eqnarray*} 
where $q_0=q(0^+)$, $g_0=g(0^+)$, $h_0=h(0^+)$,
and, after the inversion, gives
$$
\cD^\beta q = \int_0^t g(t-\tau) \cD^\beta h \, d\tau
+ h_0 J^{1-\beta}g - q_0 \frac{t^{(1-\beta)-1}}{\Gamma(1-\beta)} \,.
$$

Still about the Laplace transform,
we remind here the properties
$\mL\{\e^{-yt}g(t),s\}=\widetilde{g}(s+y)$ and
$\mL\{\e^{-yt}\int_0^t g(\zeta) \, d\zeta,s\}=
\widetilde{g}(s+y)/(s+y)$. 

To conclude, let $\sD^\mu$ be the Riesz--Feller space-fractional derivative
of order $\mu>0$ such that 
$\mF\{\sD^\mu u,\kappa\}=-|\kappa|^\mu \widehat{u}(\kappa)$
and it holds $\sD^1 \ne D^1_x$ and $\sD^2 = D^2_x$
\cite{mainardi_etal-fcaa-2001}.

\end{document}